\documentstyle[emulateapj]{article}
\begin{document}

\title{ARE CEN A AND M87 TeV GAMMA-RAY SOURCES?}

\author{J.M. Bai\altaffilmark{1,2} and Myung Gyoon Lee\altaffilmark{1}}

\altaffiltext{1}{Astronomy Program, SEES, Seoul National University, Seoul 151-742, Korea;
(jmmbai@astro.snu.ac.kr; mglee@astrog.snu.ac.kr)}
\altaffiltext{2}{Yunnan Astronomical Observatory, CAS, Kunming 650011;
National Astronomical Observatories, PR China}

\begin{abstract}
In this paper, we identify Cen A and M87, two nearby Fanaroff-Riley I (FRI)
radio galaxies, with high-energy-peaked BL Lac-like (HBL-like) objects
by investigating their spectral energy distributions (SEDs). The SED peak
of Cen A at $\sim$150 keV, which was generally believed to be the peak of
inverse-Compton emission as in the case of 3C 273, is found to be actually
the peak of synchrotron emission.
The synchrotron emission of M87 peaks in the far UV band.
We summarize the properties of $\gamma$-ray-loud blazars, especially those
of TeV BL Lac objects, and generalize them to HBL-like FRI radio galaxies
according to the unified scheme of BL Lac objects and FRI radio galaxies.
We infer that Cen A may have a peak in its Compton component power output at
$\sim$1 TeV, and that M87 may have a Compton emission peak at $\sim$0.1 TeV.
For Cen A, the estimated TeV $\gamma$-ray flux during outburst is
F(0.25 $-$ 30 TeV) = 6.4$\times10^{-9}$ erg cm$^{-2}$s$^{-1}$, and for M87,
F(0.25 $-$ 30 TeV) = 1.1$\times10^{-11}$ erg cm$^{-2}$s$^{-1}$.
Both fluxes are detectable by TeV detectors available
today, and hence Cen A and M87 are TeV $\gamma$-ray source candidates.
By investigating the long term variability, we predict that Cen A will
undergo an outburst in the near future and will be detectable at the TeV $\gamma$-ray
energy range using the CANGAROO and the German-French-Italian experiment HESS TeV
$\gamma$-ray telescopes.

\end{abstract}


\section{INTRODUCTION}

Up to now, five active galactic nuclei (AGNs) in total have been discovered
to be TeV ($E>$0.3 TeV) $\gamma$-ray sources. All of them are nearby
BL Lac objects, namely Mrk 421 (z=0.031, Punch et al. 1992), Mrk 501 (z=0.034,
Quinn et al. 1996), 1ES 2344+514 (z=0.044, Catanese et al. 1998),
PKS $2155-304$ (z=0.117, Chadwick et al. 1999),
and 3C 66A (z=0.444, Nespher et al. 1998).
BL Lac objects and flat
spectrum radio quasars (FSRQs) constitute a rare and extreme blazar class
of AGN which is characterized by strong and rapid
variability, high polarization, and apparent superluminal motion.
These extreme properties are generally interpreted
as a consequence of nonthermal emission from a relativistic jet oriented
close to the line of sight (Blandford \& K\"onigl 1979).
According to unified schemes, BL Lac objects are intrinsically the same as
Fanaroff-Riley class I (FRI, Fanaroff \& Riley 1974) radio galaxies, while
FSRQs are the same as FRII radio galaxies (e.g. Urry \& Padovani 1995).

Besides the above five TeV $\gamma$-ray-loud blazars, 66 blazars
have been 
detected to date as GeV $\gamma$-ray sources by the EGRET experiment on board
{\it CGRO} (Mukherjee et al. 1997; Mattox et al. 1997; Hartman et al. 1999;
Sreekumar et al. 1999). The $\gamma$-ray emission of blazars indicates
a double-peak structure in the overall spectral energy distribution (SED)
(Ulrich et al. 1997 and references therein; Fossati et al. 1998),
suggesting two broad spectral
components. The first component is generally interpreted as being due to
synchrotron emission, and the second is believed to be inverse-Compton
emission of the same electron population (e.g. Ulrich et al. 1997; Urry 1999).
According to the different peak frequencies, BL Lac objects are divided into
two subclasses, namely the low-energy-peaked
BL Lac objects (LBL) which have synchrotron peaks in IR/optical, and
high-energy-peaked BL Lac objects (HBL) which have synchrotron peaks in
the UV/soft X-ray region.
All of the above TeV $\gamma$-ray blazars are HBLs (the radio-selected 
BL Lac object 3C 66A has an $\alpha_{rx}=0.74<0.80$, Fossati et al. 1998)
while almost all GeV $\gamma$-ray sources are LBLs and FSRQs.

Although TeV $\gamma$-ray observations have already significantly affected 
our understanding of BL Lac objects and the unified scheme (e.g. Catanese \&
Weekes 1999),
the sample of TeV blazars is still small.
Much observational and instrumental work has been devoted to searching
for new TeV blazars (Roberts et al. 1999; Urry 1999;
Catanese \& Weekes 1999 and references therein; Aharonian et al. 2000).
As Bai \& Lee (2001) pointed out, according to the unified scheme of FRI
radio galaxies and BL Lac objects, the SED of the jet-related
emission of FRIs should also exhibit double-peak structure and some of them
should be HBL-like.
Though the relativistic boosting is week or even negative in FRIs,
some nearby HBL-like FRIs may be loud enough to be
detectable in the TeV energy range.
If this turns out to be true, it will not only enlarge the sample of TeV
sources but also provide a new approach for the study of radio galaxies.
Moreover, TeV observations of nearby FRIs will provide a test of the unified scheme.

Cen A (NGC 5128) is the prototype of FRI radio galaxies, at a distance
of 3.4 Mpc (z=0.0008) with two strong X-ray jets (e.g. Kraft et al. 2000).
It is the only GeV $\gamma$-ray source not belonging to the blazar class
(Sreekumar et al. 1999).
M87 (3C 274, Virgo A, NGC 4486, z=0.0043) is a well studied FRI radio galaxy
with the brightest optical jet. It shows most of the characteristics of BL Lac
objects, with the jet oriented at  $30\arcdeg - 35\arcdeg$ to our
line of sight (Tsvetanov et al. 1998 and references therein).

In this paper we present evidence showing that Cen A and M87 are HBL-like FRIs and
that both of them are probably loud enough to be detectable at the TeV energy
range during outburst.

\section{PROPERTIES OF $\gamma$-RAY-LOUD BL LAC OBJECTS}

The $\gamma$-ray-loud BL Lac objects (FSRQs similar to LBLs) have some
common properties which imply physical similarities among different
objects. It is believed that $\gamma$-ray emission from TeV BL Lac objects is
dominated by the synchrotron self-Compton (SSC) process.
Within the present uncertainties, the low frequency side
of each of the two peaks of TeV BL Lacs can be described by
the same spectral index (Tavecchio et al.
1998), consistent with the SSC model.
Besides the property of double-peaked SEDs
mentioned in the Introduction section,
$\gamma$-ray-loud blazars have the following properties
which should also be shared by HBL-like FRI radio galaxies,
according to the unified scheme:

\begin{flushleft}
1) The relation between the two peak frequencies
\end{flushleft}

In the context of the SSC model, the relation between the peak frequencies
of the
synchrotron component $\nu_{s}$ and the Compton component $\nu_{c}$ is
$\nu_{c}/\nu_{s} \propto \gamma_{peak}^{2}$, where $\gamma_{peak}$ is the
characteristic electron energy (Urry 1999).
For TeV BL Lac objects, the averaged upshifting factor of
the Compton peak relative to the synchrotron peak is $\sim10^{8\pm1}$, i.e.,
\begin{equation}
\nu_{c}/\nu_{s} \simeq 10^{8\pm1}.
\end{equation}
For example, Mrk 421 has a synchrotron emission peak at or slightly above
$10^{17}$Hz, and a Compton peak just below 1 TeV (Urry 1999), yielding
$\nu_{c}/\nu_{s}\lesssim10^{9}$.
The synchrotron peak of Mrk 501 occurs in the range 
2 $-$ 100keV (Catanese et al. 1997; Pian et al. 1998; Urry 1999; Catanese \&
Sambruna 2000), and the Compton component peaks at
$\sim$1 TeV (Petry et al. 2000; Sambruna 1999),
with $\nu_{c}/\nu_{s}\sim10^{7}$ -- $10^{9}$.
PKS $2155-304$ has a peak in its synchrotron component in the soft X-ray range
(Edelson et al. 1995; Urry et al. 1997; Chiappetti et al. 1999) and a peak in its Compton
component between 10GeV and 1 TeV (see Fig. 5 of Chadwick et al. 1999),
probably at 100GeV (Kataoka et al. 2000; Chiappetti et al. 1999)
with $\nu_{c}/\nu_{s}\sim10^{8}$--$10^{9}$.

Since the flux sensitivity of TeV $\gamma$-ray detectors is not high, it is necessary
for a candidate TeV source to have a peak in its Compton component in the TeV
$\gamma$-ray energy range in order for it to be detected.
BL Lac (2200+42, z=0.069) is one of the brightest LBLs.
It has a Compton radiation peak in the GeV energy range (Fossati et al. 1998)
and is a strong GeV $\gamma$-ray source
but it was not detected in the TeV energy range even during a large optical/GeV outburst in July/August 1997
(Bloom et al. 1997; Bai et al. 1999;  Aharonian et al. 2000). Thus,
TeV candidates should be selected among HBLs or HBL-like objects,
according to Eq. (1).

\begin{flushleft}
2) The relation between the two peak fluxes
\end{flushleft}

The two radiation components of blazars are self-similar in shape.
In the context of the SSC model, the luminosities
($L_{s}$, $L_{c}$), peak luminosities ($L_{s}(\nu_{s})$ and $L_{c}(\nu_{c})$),
and peak flux densities ($f_{s}(\nu_{s})$ and $f_{c}(\nu_{c})$) of the two
components are related by (Tavecchio et al. 1998; Stecker et al. 1996)
\begin{equation}
\frac{L_{c}}{L_{s}} = \frac{\nu_{c}L_{c}(\nu_{c})}{\nu_{s}L_{s}(\nu_{s})} =
\frac{\nu_{c}f_{c}(\nu_{c})}{\nu_{s}f_{s}(\nu_{s})}.
\end{equation}
During the high state, the five TeV blazars all have a Compton luminosity
$L_{c}$ comparable to or slightly less than the synchrotron luminosity $L_{s}$,
i.e., $L_{c}/L_{s} \sim $1.
Assuming this is also valid for candidate TeV sources, we obtain
\begin{equation}
\nu_{c}f_{c}(\nu_{c}) \simeq \nu_{s}f_{s}(\nu_{s}).
\end{equation}
According to the SSC model, the Compton component has the same spectral shape as the
synchrotron component. Thus, with Eq. (3) and the spectral index of the
synchrotron component, the TeV $\gamma$-ray flux of a candidate can be estimated
as
\begin{equation}
F_{\gamma} \simeq \nu_{s}f_{s}(\nu_{s})
[\frac{\nu_{c}^{1-\alpha_1}-\nu_{1}^{1-\alpha_1}}
{(1-\alpha_1)\nu_{c}^{1-\alpha_1}}+
\frac{\nu_{2}^{1-\alpha_2}-\nu_{c}^{1-\alpha_2}}
{(1-\alpha_2)\nu_{c}^{1-\alpha_2}}],
\end{equation}
where $\nu_{1}$ ($nu_{1}\leq\nu_{c}$) and $\nu_{2}$ ($\nu_{2}\geq\nu_{c}$)
are the energy thresholds of a TeV
$\gamma$-ray detector, and $\alpha_1$ and $\alpha_2$ are
spectral indices of the synchrotron component below and above $\nu_{s}$,
respectively.

\begin{flushleft}
3) The variability
\end{flushleft}

Perhaps the most striking property of TeV $\gamma$-ray sources is that
the TeV flux is strongly variable.
The simultaneous multiwaveband observations for
Mrk 421 and Mrk 501 show that the amplitude of variations at the TeV energies
is the
largest (Petry et al. 2000; Catanese et al. 1997; Takahashi et al. 1996).
Both of them exhibited low states in which the TeV flux is below the limit of
detection.
It is thus obvious that during the low state $L_{c}/L_{s} < $1, and sometimes
$L_{c}/L_{s} << $1.
For this reason, Eqs. (3) and (4) are only valid during outburst.

For Mrk 421 and Mrk 501, TeV flux varies with X-rays without a significant time
lag ($<$24hr).
This may be caused by changes in electron distribution, so it is likely that other
TeV AGNs may also show similar behavior.
Therefore, the TeV $\gamma$-ray outburst of an object can
be predicted by investigating the X-ray variability of the object.

\begin{flushleft}
4) Beaming effects
\end{flushleft}

For $\gamma$-ray-loud blazars, relativistic beaming is necessary
not only to enable the
$\gamma$-ray photons to escape from the source, but also to amplify the flux
and therefore make the source more easily detectable (e.g. Urry \& Padovani
1994; Ghisellini et al. 1998).
The observed ($F_{obs}$) and intrinsic
($F_{0}$) fluxes from a continuous relativistic jet are related by
\begin{equation}
F_{obs} = \delta^{2+\alpha}F_{0} = \delta^{p}F_{0},
\end{equation}
where $\delta$ is the Doppler factor and $\alpha$ is the spectral index.
The Doppler factor is defined as
$\delta = [\Gamma(1-\beta\cos\theta)]^{-1}$, where $\beta$ is the speed
of emitting plasma in units of the speed of light, $\Gamma$ is the Lorentz
factor, defined as $\Gamma=(1-\beta^2)^{-1/2}$, 
and $\theta$ is the viewing angle.
It is notable that beaming does not affect the relations between two
components discussed above (see Fig. 5 in Chiaberge et al. 2000).

Based on the TeV variability, it was estimated that Mrk 421 and Mrk 501 have
$\theta\lesssim6\arcdeg$ (Padovani 1999).
If they were observed at $\theta=60\arcdeg$,
according to the Eq. (5) and
assuming typical values for $\Gamma$ and $p$, i.e., $\Gamma\sim$5 and $p\sim$3,
the observed flux would decrease
by a factor of $1.3\times10^{-4}$, which is not detectable even during the
largest outburst. However, if they were at the distance of Cen A, z=0.0008,
the observed flux would increase by a factor of about
$(\frac{0.034}{0.0008})^2$.
Therefore, if Mrk 421 and Mrk 501 were moved to the position of Cen A and
observed at a
viewing angle of $60\arcdeg$, or moved to the position of M87 (z=0.0043)
and observed at $\theta=35\arcdeg$, they would still be detectable at 
TeV energies during outburst.

\section{CEN A AND M87 AS TeV SOURCE CANDIDATES}

As mentioned in Section 1,
BL Lac objects are divided into HBLs and LBLs according to the peak
locations in the SEDs. Further studies show that
for LBLs the X-rays are inverse-Compton emission, while X-rays in HBLs are an
extension of synchrotron emission at lower energies (Padovani \& Giommi 1996;
Fossati et al. 1998). In the following subsections we will show that X-rays
in Cen A and M87 consist of synchrotron emission and thus they are HBLs.

\subsection{Cen A}

Figure 1 shows spectral energy distributions of Cen A, M87 and Mrk 501
based on several observations in the literature.
The X-ray spectral index of Cen A nucleus in 2-10 keV was found to be less
than 1 (Turner et al. 1997; Miyazaki et al. 1996), and hence
$\log(\nu f_{\nu})$ is a monotonously rising function of $\log\nu$.
Between 10 and 100 keV, the spectral index is 0.7 $-$ 0.8 (Miyazaki et al.
1996; Bond et al. 1996; Wheaton et al. 1996; Kinzer et al. 1995), also less
than 1, with a monotonously rising SED of Cen A in this energy range.
The SED of Cen A continuously rises below $\sim$ 150 keV where
the spectrum breaks, then the SED goes down continuously to the MeV and
even GeV energy range (see Fig. 1).
As in the case of blazars, this hump represents one of the two radiation
components, and the radiation mechanism of this component is the same as that
of X-rays.

Steinle et al. (1998) and Kinzer et al. (1995) pointed out that the spectra
of Cen A show interesting
similarities to those of jet-aligned blazars (McNaron-Brown et al. 1995), and in
particular to those of the well-studied quasar 3C 273 (Johnson et al. 1995;
Lichti et al. 1995); both show spectral breaks in the soft $\gamma$-ray
regime, and both have intensity-independent power-law shapes below the
break. For 3C 273 the peak in the soft $\gamma$-ray regime is a Compton peak,
so Steinle et al. (1998) and Kinzer et al. (1995) regarded the peak of Cen A
at $\sim$150 keV as a Compton peak.

However, based on the observations by {\it ROSAT} and {\it ASCA}, Turner et al.
(1997) found that the spectrum of the jet of Cen A is actually
consistent with the spectrum of the nuclear continuum,
and that the spectrum of the jet is
flatter than that expected as a result of an inverse-Compton scattering
of radio photons but is consistent with that predicted by a simple
synchrotron model. Recent Chandra observations also indicated that X-rays
from the jet of Cen A comprise synchrotron emission (Birk \& Lesch 2000).
Therefore, the peak at $\sim$150 keV is not the Compton peak but
the synchrotron peak, as the case of Mrk 501 (see Fig. 1).
The peak frequency of Cen A is variable, which is also similar to the behavior in TeV blazars.
Observations by {\it CGRO} OSSE in 1991 $-$1994 revealed peak frequencies of
$\sim$150 keV (Kinzer et al. 1995),
while Miyazaki et al. (1996) reported a peak
frequency of 180 keV, based on Welcome-1 observations.

The peak of the Compton component of Mrk 501 is at $\sim$1 TeV,
so that of Cen A is probably also at $\sim$1 TeV.
During the outburst state, the flux density at 100 keV is
$\approx 11\times10^{-5}$ photon cm$^{-2}$s$^{-1}$keV$^{-1}$ (see Fig. 5 in
Bond et al. 1996), and the synchrotron peak flux density is $f_{150keV} =
2.1\times10^{-29}$erg cm$^{-2}$s$^{-1}$Hz$^{-1}$, assuming a spectral index
of 0.7. According to Eq. (4) with $\alpha_1$=0.7 and $\alpha_2$=1.3,
the estimated TeV flux in 0.25 $-$ 30 TeV for Cen A is
\begin{equation}
F(0.25 - 30 {\rm TeV}) = 6.4\times10^{-9} {\rm erg}\, {\rm cm}^{-2}{\rm s}^{-1}
\end{equation}
which is about 10 times brighter than that of Mrk 501 in the 1997 outburst,
suggesting that Cen A can be a strong TeV $\gamma$-ray source during outburst.

\subsection{M87}

The radio to X-ray spectral energy distribution of M87 is very smooth
like that of a typical BL Lac object
(Tsvetanov et al. 1998; see Fig. 1).
If it is really a misaligned BL Lac object, the radio to
X-ray hump should be the synchrotron component, and X-rays are also synchrotron
emission, suggesting that M87 is an HBL.
{\it EINSTEIN} and {\it ROSAT} observations show that the X-ray emission from the jet
of M87 is probably the high frequency tail of the radio to optical synchrotron
spectrum, although other origins (thermal or inverse-Compton) for the X-rays
cannot be ruled out completely (Biretta, Stern \& Harris 1991;  Neumann et
al. 1997).
The $EUVE$ observations of M87 are consistent with the spectral cutoff in the
spectrum of the jet in M87 as suggested by Meisenheimer et al. (1996), which
further supports the idea that the EUV and X-ray emission of the jet is
synchrotron radiation (Bergh\"ofer et al. 2000).
Reynolds et al. (1999) also identify M87 with an HBL, by
combining {\it Rossi X-ray Timing Explorer (RXTE)} hard X-ray (4 $-$ 15keV)
observations with {\it ROSAT} data and inferring that the X-ray
spectra of the M87 core and jet must be steep like those of HBLs.

The synchrotron peak of M87 lies between $10^{15} - 10^{16}$ Hz (see Fig. 1).
According to Eq.(1), the Compton component of M87 may peak at $\sim$0.1 TeV.
The flux of the core and jet of M87 in the $EUVE$ DS band pass is 16.25
$\mu$Jy (using a core/jet flux ratio of $\sim$1.5)
with a spectral index of the UV to X-ray power-law spectrum of
$\alpha \approx 1.4$ (Bergh\"ofer et al. 2000; Harris et al. 1997).
According to Eq. (4), the estimated TeV flux (0.25 $-$ 30 TeV) for M87 is
\begin{equation}
F(0.25 - 30 {\rm TeV}) = 1.1\times10^{-11} {\rm erg}\, {\rm cm}^{-2}{\rm s}^{-1}
\end{equation}
which is comparable to that of Mrk 501 in 1995,
suggesting that M87 may be a TeV $\gamma$-ray-loud FRI radio galaxy.

\section{DISCUSSION AND CONCLUSIONS}

Cen A was observed with the CANGAROO 3.8m TeV $\gamma$-ray telescope from March
to April 1995, but was not detected (Rowell et al. 1999).
There was no X-ray observation for Cen A in 1995, but using the observations
before 1995 we can infer that Cen A was in the low state during 1995.
Between 1992-1995 Cen A was very faint and there was no trend towards outburst in 1995
(see Fig. 5 in Bond et al. 1996). Thus, the TeV flux of Cen A was
below the detection limit of the CANGAROO telescope.
Grindlay et al. (1975) reported that Cen A was detected in the TeV
energy range during 1972-1974.
Since the detection has not been repeated (Israel 1998), it
appears that no one has included Cen A in the list of known TeV $\gamma$-ray AGNs.
However, from 1972 to 1976, Cen A underwent a large outburst in the X-ray
energy range (see Fig. 4 in
Turner et al. 1997 or Fig. 5 in Bond et al. 1996), and probably
underwent an outburst 
in the TeV energy range at the same time, so the detection of Cen A by
Grindlay et al. (1975) is considered to be authentic.

At present, the CANGAROO 10m telescope is the only TeV $\gamma$-ray telescopes
which can be used to observe the southern object Cen A.
In the future, Cen A can be observed by CANGAROO III
and the German-French-Italian experiment HESS in Namibia.
The long-term X-ray light curve of Cen A shows that the period of outburst
is $5 - 6$ years (see Fig. 5 in Bond et al. 1996), which is typical for
radio-loud AGNs (Smith and Nair 1995).
The recent observation by the Chandra X-ray Observatory in 1999 shows that
Cen A is in the low state at present (Kraft et al. 2000).
Thus, it is expected that Cen A may undergo an outburst in the
near future.
The long-term light curve also shows that Cen A did not undergo any large
outburst since 1986, so the coming outburst may be a large one.
In addition, according to Eq. (1) the Compton component of Cen A may also
peak as high as $\sim$100 TeV.
It may thus be strong enough to be detectable
even in the PeV energy range during outburst.

The {\it HST} has recorded that the flux
from the M87 optical nucleus varies by a factor of $\sim$2 on timescales of
$\sim$2.5 months (Tsvetanov et al. 1998).
A difference by a factor of 
five in the flux detected by {\it Spacelab 2} and Ginga indicated that
the hard X-ray from M87 core also varied violently (Guainazzi \& Molendi 1999),
which suggests that the variability in M87 around peak frequency is similar to those
in TeV blazars.
EGRET has observed M87 but did not detect any GeV
$\gamma$-rays (Sreekumar et al. 1996),
which may indicates that M87 does not have a peak in its Compton component at the GeV energy
range, being consistent with our result.
M87 was observed at TeV energies many years ago but was not detected.
At that time the TeV detectors
were not sensitive (the Crab Nebula was not detected either), and the observations only
gave upper limits, 
$F(>0.4 {\rm TeV}) < 8.3\times10^{-11} {\rm photons}\, {\rm cm}^{-2}{\rm s}^{-1}$
(Cawley et al. 1985) and
$F(>0.21 {\rm TeV}) < 1.2\times10^{-10} {\rm photons}\, {\rm cm}^{-2}{\rm s}^{-1}$
(Weekes et al. 1972; see Fig. 1).
M87 has not yet been observed by any modern TeV $\gamma$-ray telescopes.
It can be observed using powerful TeV $\gamma$-ray telescopes, such as Whipple and HEGRA.

In summary,
the X-rays from Cen A and M87 are synchrotron emission rather than
inverse-Compton emission, and hence
these nearby FRI radio galaxies are HBL-like objects.
In particular, Cen A is similar to Mrk 501 with its synchrotron component peaking
at very high energies, which was once thought to be non-synchrotron radiation
(Skibo et al. 1994; Kinzer et al. 1995; Steinle et al. 1998).
According to the unified scheme of BL Lac objects and FRI radio galaxies
and assuming that the properties of the known TeV BL Lac objects are common
for all TeV AGNs, we predict that Cen A
may have a peak in its Compton component power output at $\sim$1 TeV,
and that M87 may have a Compton emission peak at $\sim$0.1 TeV,
both having TeV $\gamma$-ray flux detectable by TeV $\gamma$-ray
detectors available today.

\acknowledgments

This work was financially supported by the BK21 Project
of the Korean government. 
We thank Helmut Steinle and Tracey Jane Turner for 
providing us with soft $\gamma$-ray and X-ray data, respectively.
We also thank the anonymous referee for helpful suggestions.


\figcaption{Spectral energy distribution of Cen A (filled circles), M87 (asterisks)
and Mrk 501 (open circles).
For Cen A, the data from radio to optical bands are from Marconi et al. (2000),
and X-ray and soft gamma-ray data obtained in 1991 are from Steinle et al.
(1998).
For M87, EUV data are from
Bergh\"ofer et al. (2000), UV data are from Sparks et al. (1996), and
other data are from Biretta et al. (1991).
For Mrk 501, the SED is based on simultaneous observation of MJD=50556.32
(Petry et al. 2000).
The upper arrow represents upper limit from Weekes et al. (1972), and the lower
arrow represents upper limit from Cawley et al. (1985).}

\end{document}